\documentclass[prb,twocolumn,showpacs,floatfix]{revtex4}

\usepackage{graphicx}
\usepackage{amsmath,amssymb,bm,dcolumn}
\usepackage[export]{adjustbox}
\usepackage{graphicx}
\usepackage{float}
\usepackage{amsmath}
\usepackage{amsfonts}
\usepackage{color}
\usepackage{slashed}
\usepackage[colorlinks,hyperindex]{hyperref}

\hypersetup{
colorlinks,
citecolor=blue,
linkcolor=green,
urlcolor=black,
}

\def\be{\begin{equation}}
\def\ee{\end{equation}}
\def\bed{\begin{description}}
\def\eed{\end{description}}
\def\bea{\begin{eqnarray}}
\def\eea{\end{eqnarray}}

\def\ba{\begin{array}}
\def\ea{\end{array}}

\def\u1{$U(1)$}
\def\suu1{$SU(2)\times U(1)$}

\begin{document}

\title{Lorentz Violation and Topologically Trapped Charge Carriers in 2D Materials}
\author{R. A. C. Correa$^{1}$, W. de Paula$^{1}$, A. de Souza Dutra$^{2}$,
T. Frederico$^{1}$}
\affiliation{
 $^1$ Instituto Tecnol\'{o}gico de Aeron\'{a}utica, DCTA, 12228-900, S\~{a}o Jos\'{e} dos Campos, SP, Brazil \\
$^2$DFQ, Universidade Estadual Paulista-UNESP, 12516-410, Guaratinguet\'{a},
SP, Brazil}

\begin{abstract}
The full spectrum of two-dimensional fermion states in a scalar soliton trap with a Lorentz breaking background is investigated in the context of the novel 2D materials, where
the Lorentz symmetry should not be strictly valid. The field theoretical model with Lorentz breaking terms represents  Dirac electrons in one valley and in a scalar 
field background. The Lorentz violation comes from the difference 
between the Dirac electron and scalar mode velocities, which should be expected when modelling the electronic and lattice excitations in 2D materials. 
We extend the analytical methods developed in 
the context of 1+1 field theories to explore the effect of the Lorentz symmetry breaking in the  charge carrier density of 2D materials in the presence of a domain wall with a  kink profile. The
width and the depth of the trapping potential from the kink is controlled by the 
Lorentz violating term, which is reflected analytically in the band structure and 
properties of the trapped states.  Our findings enlarge
previous studies  of the edge states obtained with domain wall and  in strained graphene nanoribbon in a chiral gauge theory.

%

\end{abstract}

\maketitle

\section{Introduction}

Since the seminal work by Skyrme \cite{Skyrme-PRSA-1961}, where the first
three-dimensional topological defect solution arising from a nonlinear field
theory was presented in the context of particle physics, a very large number
of studies reporting the impact of some kind of topological defect has
appeared in the physics literature and, in recent years, this area of
research remains one the most active fields in many areas of the physics,
including condensed matter physics \cite{Topological-CM}, field theory \cite%
{Topological-FT}, and cosmology \cite{Topological-CO}. For instance, in an
astrophysical scenario, topological defects arises largely from grand
unified theories (GUTs) of elementary particles. In this case, it is
expected that the known local gauge symmetry group $SU_{C}(3)\times
SU_{L}(2)\times U_{Y}(1)$\ resulted from an underlying symmetry group $G$\
after a series of spontaneous symmetry breakings. Thus, the early universe
has gone through a number of phase transitions, with one or more several
types of topological defects possibly being left behind.

Another interesting background, where we can find topological
configurations, is high energy physics \cite{Rajaraman-Solitons}. In this
context, topological structures are found for example in the massive
Thirring model \cite{Orfanidis}, in nonlinear chiral theories \cite%
{Duff-Isham}, in the Gross-Neveu model \cite{Klein}, in a chiral $SU(3)$\ $%
\times $\ $SU(3)$\ model \cite{Nicole}, in an artificial spin-ice lattice 
\cite{Sebastian}, and where the Peccei-Quinn symmetry is broken after
inflation \cite{Kawasaki}.

Nowadays we know that the carbon is the key ingredient in the called organic
chemistry. From the viewpoint of condensed matter physics, the carbon plays
a crucial role in the formation of new classes of topological materials \cite%
{CasPRP09} with a large variety of physical properties \cite{Peres,VozPRP10}%
. In particular, the material named in the literature as graphene is the
most famous due to their great potential for technological applications.
Those materials are a two-dimensional (2D) allotrope of carbon, which
consists of a monotonic layer of carbon atoms on a honeycomb lattice. Its
abundant electronic properties was studied for the first time using the
tight-binding approximation \cite{Wallace}, where the structure of the
electronic energy bands and Brillouin zones for graphite has been
calculated. Although it is a nonrelativistic system, because the fermions
move with a speed $v_{F}$, which is 300 times smaller than the speed of
light $c$, the most remarkable feature of the graphene is that electronic properties
can be described by the relativistic Dirac equation, as initially pointed by
Semenoff \cite{Semenoff}. Thus, we can use the well-established methods of
quantum field theory in order to study the physics and topological
properties of the graphene\cite{Chaves:2010zn, Chaves:2013fca}.

A special point of interest in graphene are the topological excitations,
which are a direct consequence of its strong chemical bonding \cite{Pachos}
responsible for assuring a stability in the geometrical properties of the
underling lattice. Topological defects in graphene can lead to the
appearance of an effective gauge field, which can be induced by wrinkles or
ripples in its structures \cite{Guinea}, when the graphene sheet is under
tension \cite{Prada}, and by displacements along bond directions \cite%
{Chamon}.

{\ Topological excitations of the 2D lattice have its counterpart in the
electronic properties of the material. One example, is the Jackiw-Pi theory
that predicts a Dirac zero mode, associated with vortex solutions of the
gauge field coupled to the electrons, leading to charge fractionalization 
\cite{JacPRL07, ChamPRB08}. The Jackiw-Pi theory was further generalized to
non-chiral models \cite{OlivPRB11, Popovici:2012xs} , where the implications of different
gauge groups in the gap formation, considering scalar, vector, and fermion
excitations were studied. In addition the finite-energy vortex solutions of
the gauge models having the flux of the electromagnetic fields quantized
leads to the Bohm-Aharonov effect even in the absence of external
electromagnetic field. }Furthermore, we can find an interesting
supersymmetric model for graphene in Refs. \cite{EvertonJHEP2011,
EvertonAOP2015}.

Another example, of interest, is the possibility of solitonic excitations in
the 2D lattice from the nonlinear solutions of the field equations in
nanoribboins (see e.g. \cite{CorPRB13}). The static deformation of the
scalar field forming a domain wall in a graphene layer has been analyzed in 
\cite{SemPRL08}, where it was discovered an electronic domain wall spectrum
associated with localized electronic states transverse to the wall, and
progating only in one dimension. They studied gapped graphene and found
three confined states, one the zero-mode and one in the conducting and
valence band. In strained graphene nanoribbon it was also studied a soliton
trap which has electronic states confined to the wall associated with a
Dirac zero mode solution in the trap \cite{SasNJP10}.

The topological confinement of electrons in bilayer graphene was proposed in 
\cite{MarPRL08}, where also the associated spectrum was computed numerically
(for a recent review on the electronic properties of bilayer graphene see 
\cite{RozPRP16}). Such theoretical proposal was experimentally verified by
observing the valley transport properties at the bilayer graphene domain
wall \cite{JuNAT15}, and quite recently \cite{LiNATNAN16} it was possible to
control these one-dimensional conducting channels in a gapped set up. The
imaging of the confined states was done in \cite{YinNATC16} by using
scanning tunnelling microscopic, and it was found that the associated
conducting channels are mainly located close to the two edges of the domain
wall.

We should remind that Jackiw and Rebbi \cite{RebbiPRD1976} in 1976 proposed
that Fermions can couple to bosonic soliton configurations in non-linear
theories in (1+1) dimensions. In that work, it was concluded that there is a
zero-energy bound state, where the solitons are degenerate doublets with
fermion number $\pm 1/2$. Following this seminal paper, several works have
been done with theories involving fermionic fields coupled to topological
defects, such studies have included fermions and vortex solutions in abelian
and nonabelian gauge theories \cite{VegaPRD1978}, monopoles \cite%
{PaoloNPB1977}, supersymmetric field theories \cite{VecchiaNPB1977},
condensed matter and relativistic field theories \cite{JackiwNPB1981},
nanorings \cite{IgorPRB2013}, dense quark matter \cite{EtoPTEP2014},
Aharonov-Bohm and Aharonov-Casher problems \cite{KhalilovEPJC2014}, and
topological superconductors \cite{SahooPRB2016}.

Motivated by the work of Ref. \cite{RebbiPRD1976}, Chu and Vachaspati \cite%
{ChuPRD2008} obtained the full spectrum of fermion bound states on a scalar Z%
$_{2}$ kink in (1+1) dimensions. Indeed, a decade before in \cite%
{DavisPRD1988}, it was speculated that the presence of possible fermionic
bound states would inhibits the decay of cosmic strings. 
The method developed in Ref. \cite{ChuPRD2008} to solve the Dirac equation
on a kink background has been shown to be extremely powerful. It enables to
obtain analytically the bound state solutions. Later on, Dutra and Correa 
\cite{RafaelPLB2010} have extended such method to study fermion bound-states
and zero modes in the background of kinks within models presenting a
structure with a false and a true vacuum. The spectrum of fermion
bound-states of a class of asymmetrical scalar field potentials was
analytically obtained in that work. Therefore, the findings of \cite%
{ChuPRD2008,RafaelPLB2010}, in one-dimensional theory, suggest that a domain
wall, or a soliton trap for electrons in 2D materials, should also presents
an analytical representation of the rich spectrum and the corresponding
states adding this feature to the previous finding of \cite{SemPRL08} and 
\cite{SasNJP10}.

In the present work we generalize the analytical methods developed in \cite%
{ChuPRD2008} for field theories of fermions coupled to scalar kinks in (1+1)
to (2+1) dimensions in order to obtain the electronic spectrum in the
presence of such topological defect. We found analytically the excited
states trapped by this domain wall, adding those to the ones previously
obtained by \cite{SemPRL08} and \cite{SasNJP10}. Furthermore, we will
explore the properties of the spectrum under the influence of constant
external fields coupled to the fermions and also the associated impact due
to the change of the kinetic term of the scalar field, which is given by Lorentz symmetry violation terms in the Lagrangian. Our motivation is to
check the robustness of the electron spectrum when the properties of scalar
field associated with modes of the 2D lattice are modified, in particular
with respect to the propagation speed.

In fact, in a pioneer work, Kostelecky and Samuel \cite{LV4}
proposed Lorentz Symmetry Violation (LSV) in string theories, that was
explored in several other physical scenarios (see e.g. \cite{KosRMP2011} and
references therein). In particular, some works have explored topological
defects in the presence of LSV. For instance, we can find investigations
using monopole and vortices \cite{LV38, LV39,LV40,LV41,LV42,LV43}, in a
gauged $O(3)$\ sigma model \cite{LV44}, on topological defects generated by
two real scalar fields \cite{LV45a,LV45b,LV45c}, in the propagation of
electromagnetic waves \cite{LV46}, and in traveling solitons systems \cite%
{LV48, LV51}.

This paper is organized as follows: in Section II we present the theoretical
framework of the SME which is going to be analyzed and we find the classical
field configurations associated to it. In Section III we compute the full
spectrum of two-dimensional fermion states. In Section IV we present our
conclusions and final remarks.

\section{Extended Lagrangian with Lorentz-violating interaction}

\label{setup}

In this section, we will show the general form of the relativistic
Lagrangian for a free spin-$1/2$ Dirac fermion $\psi $ of mass $m$ in the
so-called standard-model extension. At this point it is important to remark
that this theory is based on the idea of spontaneous Lorentz breaking in an
underlying theory and has been used for various investigations placing
constraints on possible violations of Lorentz symmetry, several of which
depend crucially on the nonrelativistic physics of free massive fermions.

Our development is a generalization of the work \cite{ChuPRD2008}, where it
was studied the kink-fermion coupling in $1+1$ dimension without Lorentz
breaking. We extend the $1+1$ theory studied in \cite{ChuPRD2008} to $2+1$
dimensions including Lorentz breaking terms in the model in order to study
the electronic properties of 2D materials in the presence of a kink. The
extended Lagrangian density, adopted in the present work, has the following
form: 
\begin{eqnarray}
\mathcal{L} &=&\frac{1}{2}\,\partial _{\mu }\phi \partial ^{\mu }\phi
+k^{\mu \nu }\,\partial _{\mu }\phi \,\partial _{\nu }\phi -V(\phi )+i{\bar{%
\psi}}\gamma ^{\mu }\overleftrightarrow{\partial }_{\mu }\psi  \notag \\
&&-a_{\mu }{\bar{\psi}}\gamma ^{\mu }\psi -i\,b_{\mu }{\bar{\psi}}\gamma
_{5}\gamma ^{\mu }\psi -g\phi {\bar{\psi}}\psi ,  \label{lagrangian}
\end{eqnarray}

\noindent where $A\overleftrightarrow{\partial }_{\mu }B\equiv A\partial
_{\mu }B-(\partial _{\mu }A)B$, $\phi $ is a real scalar field, $\psi $ is a
two-component spinor, $V(\phi )$ is the potential, given in terms of a real
scalar field, defining the bosonic sector of the specific model under
analysis, $g$ is the corresponding Yukawa coupling , $k^{\mu \nu }$ is a
dimensionless coefficient for Lorentz violation that preserves\textit{\ CPT}%
. It can be taken as real, symmetric, and traceless. The polarization tensor 
$k^{\mu \nu }$ in the context of 2D materials, represents the change in the
scalar mode propagation with the direction, and its speed can be different
from the electron Fermi velocity.

Furthermore, the quantities $a_{\mu }$ and $b_{\mu }$ are parameters that
control the extent of Lorentz and \textit{CPT} violation in the theory. At
this point, it is important to remark that, in the context of the
standard-model and quantum electrodynamics (QED) extensions, these
parameters are determined by expectation values of Lorentz tensors arising
from spontaneous Lorentz breaking in a more fundamental theory. Finally, in
the above Lagrangian density, the $\gamma ^{\mu }$ are the Dirac matrices, which in this work will be written as 
\begin{eqnarray}
\gamma ^{0} &=&\sigma ^{3}=\left[ 
\begin{array}{cc}
1 & 0 \\ 
0 & -1%
\end{array}%
\right] \ ,\ \ \gamma ^{1}=i\sigma ^{1}=i\left[ 
\begin{array}{cc}
0 & 1 \\ 
1 & 0%
\end{array}%
\right] ,  \label{gm1} \\
\ \gamma ^{2} &=&i\sigma ^{2}=\left[ 
\begin{array}{cc}
0 & 1 \\ 
-1 & 0%
\end{array}%
\right] .
\end{eqnarray}

In order to justify the choice of such matrix representation, we point out
that some years ago, these description has enabled to obtain the spectrum of
fermion bound-states on the background of kinks of a class of asymmetrical
scalar field potentials \cite{RafaelPLB2010}. In that work, such matrix
structure allowed to find analytically the fermion configurations which has
important consequences for cosmological and condensed matter systems.
Another important matrix that will be useful is the $\gamma _{5}$. In this
context, its representation in given by 
\begin{equation}
i\,\gamma _{5}=-i\,\gamma ^{0}\gamma ^{1}\gamma ^{2}=-\left[ 
\begin{array}{cc}
1 & 0 \\ 
0 & 1%
\end{array}%
\right] .  \label{gm2}
\end{equation}
Note that for the bidimensional irreducible representation of the spinor space, the pseudovector
and vector Lorentz breaking terms in the Lagrangian are the same, as the
rotation of 180$^o$ in the plane is equivalent to the parity transformation,
and mathematically the algebra is closed with the identity and the three
Pauli matrices. Assuming isotropy for the given frame, namely $\vec a=\vec
b=0$, the net effect of the corresponding Lagrangian terms, produce only a
trivial shift the Dirac spectrum and from now on we set $a_0=b_0=0$.

Here, we would like to emphasize that, as expected, the gamma-matrices Eq. (%
\ref{gm1}) and Eq. (\ref{gm2}) satisfy the usual Clifford algebra%
\begin{eqnarray}
\{\gamma ^{\mu },\gamma ^{\nu }\} &=&\gamma ^{\mu }\gamma ^{\nu }+\gamma
^{\nu }\gamma ^{\mu }=2g^{\mu \nu },
\end{eqnarray}

In this work, as an illustrative example, we choose the $\phi ^{4}$ theory 
\cite{LV72, LV73} which is given by a symmetric double-well potential%
\begin{equation}
V(\phi )=\frac{\lambda }{4}(\phi ^{2}-\eta ^{2})^{2},
\end{equation}
where $\lambda $ and $\eta $ are real parameters. As we can see,
this double-well potential has two minima, at $\phi =\pm \eta $. Moreover,
there is reflectional symmetry since $V(\phi )=V(-\phi )$ and the vacuum
manifold has two-fold degeneracy. At this point, it is important to remember
that, in this model, the vacumm has an expectation value $<\phi >$ $=\pm
\eta $. Therefore, in such a vacuum, the elementary fermion will have a mass 
$m_{f}=\left\vert g<\phi >\right\vert =g\eta $, where we are taking $g>0$
and $<\phi >$ $=\eta $. On the other hand, the mass associated to the scalar
sector is $m_{s}=\sqrt{2\lambda }\eta $.

As we are studying a theory in $2+1$ dimensions, the tensor $k^{\mu \nu }$
in Eq. (\ref{lagrangian}) is represented by a\ $3\times 3$ matrix written in
the form 
\begin{equation}
k^{\mu \nu }=\left( 
\begin{array}{ccc}
k^{00} & k^{01} & k^{02} \\ 
k^{10} & k^{11} & k^{12} \\ 
k^{20} & k^{21} & k^{22}%
\end{array}%
\right) .
\end{equation}

It is important to remark that the above matrix has arbitrary elements.
However, if this matrix is real, symmetric, and traceless, the \textit{CPT }%
symmetry is kept. Recently, a great number of works using a similar process
to break the Lorentz symmetry, with a tensor like $k^{\mu \nu }$, have been
used in the literature, from microscope \cite{LV46} to cosmological scales 
\cite{LV51, LV60,LV36, LV37}.

From the Lagrangian density (\ref{lagrangian}), the Euler-Lagrange equation
for $\phi $ is 
\begin{equation}
\partial _{\mu }\partial ^{\mu }\phi +k^{\mu \nu }\partial _{\mu }\partial
_{\nu }\phi +V_{\phi }(\phi )=0,
\end{equation}
where $V_{\phi }(\phi )=dV(\phi )/d\phi $. Note that, in the above
equations we are assuming that the back reaction due to the Yukawa coupling
between the Dirac field and the scalar one can be neglected \cite{LV74, LV75}%
, in other words, the scalar field behaves like a classical background field 
\cite{LV76}.

As we are interested in static solutions, where the field $\phi $ has
dependence only in $z$, namely $\phi =\phi (z)$. Thus, the above equation
becomes 
\begin{equation}
\partial _{\tilde{z}}^{2}\phi =\lambda \left( \phi ^{2}-\eta ^{2}\right)
\phi ,
\end{equation}
where $\partial _{\tilde{z}}^{2}\phi \equiv d^{2}\phi /d\tilde{z}%
^{2}$, $\phi =\phi (\tilde{z})$ with $\tilde{z}=z/\sqrt{1-k^{11}}$. We
observe that the dimensionless factor $\sqrt{1-k^{11}}$ renormalizes the
propagation velocity of the scalar mode with respect to the Fermion maximum
speed, namely 
\begin{equation}  \label{velratio}
\frac{v_\phi}{v_F}=\sqrt{1-k^{11}} \, .
\end{equation}
This trivially breaks the covariance of the theory under Lorentz boosts, but
applying it to 2D materials, seems rather natural, as the propagation speed
of perturbations in the lattice is not required to be the same as the one
for electrons, which in the massless case, corresponds to the Fermi
velocity, that for graphene is $v_F\sim c/300$.

An analytic solution of the above nonlinear equation, which interpolates
between the different boundary conditions, is the so-called kink. In this
case, follows that 
\begin{equation}
\phi ({z})=\eta \tanh \left[ \frac{\alpha_0({z}-{z}_{0})}{2}\right] ,
\label{kinkphisolution}
\end{equation}
where ${z}_{0}$ is a constant of integration, it can be viewed as specifying
the position of the kink. For simplicity, and without loss of generality, we
can choose a localized kink centred about ${z}_{0}=0$, which takes $\phi $
from $-\eta $ at $z=-\infty $ to $\eta $ at $z=\infty $. The size of the
kink is controlled by: 
\begin{equation}  \label{alpha0}
\alpha _{0}\equiv \sqrt{2\lambda }\eta /\sqrt{1-k^{11}}.
\end{equation}
Note that by varying the speed of the scalar mode by the LV term in the
Lagrangian the size of the wall can be tuned.

Now, from Eq. (\ref{lagrangian}), we have the following Dirac equation in
the kink background 
\begin{equation}
\left( i\gamma ^{\mu }\partial _{\mu }-a_{\mu }\gamma ^{\mu }-i\,b_{\mu
}\gamma _{5}\gamma ^{\mu }-g\phi \right) \psi =0,  \label{dirac}
\end{equation}%
where $\phi $ is the kink solution given by the Eq.~(\ref{kinkphisolution})
written in terms of the $z$ variable.

From now on, we will solve explicitly the Eq. (\ref{dirac}). At this point,
it is important to highlight that we will only be interested in determining
the bound states. Then, as we are working with two spatial dimensions, the
fermions are described by two-component spinors, which we will write as 
\begin{equation}
\psi (y,z;t)=\frac{e^{i(\omega y-Et)}}{\sqrt{2}}\left[ 
\begin{array}{c}
\beta _{+}(z)-\beta _{-}(z) \\ 
\beta _{+}(z)+\beta _{-}(z)%
\end{array}%
\right] .  \label{sd2}
\end{equation}

Therefore, using the above representation into Eq. (\ref{dirac}), and after
some straightforward manipulations, we can find the following coupled pair
of first order differential equations 
\begin{eqnarray}
\partial _{z}\beta _{+}+g\phi \beta _{+} &=&-\left( E+\omega \right) \beta
_{-},  \label{betap} \\
\partial _{z}\beta _{-}-g\phi \beta _{-} &=&\left( E-\omega \right) \beta
_{+},  \label{betaequations}
\end{eqnarray}

From now, we will turn our attention concerning the problem of how proceed
to decouple the equations (\ref{betap}) and (\ref{betaequations}). In this
case, our goal will be to reduce the differential equations in Schr\"{o}%
dinger-like equations, which can be used with advantage in order to
understand the essential features of the quantum mechanics of system under
analysis. Then, after the substitution we obtain the corresponding second
order differential equations 
\begin{eqnarray}
-\partial _{z}^{2}\beta _{+}+V_{+}\left( \phi \right) \beta _{+}
&=&\varepsilon ^{2}\beta _{+},  \label{schrodinger} \\
-\partial _{z}^{2}\beta _{-}+V_{-}\left( \phi \right) \beta _{-}
&=&\varepsilon ^{2}\beta _{-}.  \label{schrodinger1}
\end{eqnarray}%
where we are using the corresponding definitions 
\begin{equation}
V_{\pm }(\phi )\equiv g\left[ (g\phi ^{2}\mp \partial _{z}\phi )\right] .
\end{equation}

\noindent and%
\begin{equation}
\left. \varepsilon ^{2}\equiv E^{2}-\omega ^{2}.\right.
\end{equation}

Moreover, we can note that Eq.~(\ref{schrodinger}) and Eq. (\ref%
{schrodinger1}) are one-dimensional Schr\"{o}dinger-like equations with
potential energy given by\textbf{\ }$V_{\pm }[\phi (z)]$, and where both
solutions has the same eigenvalue $\varepsilon ^{2}$.

In the next section, we will calculate both a zero mode and a fermionic
bound state on the background of kink given by Eq. (\ref{kinkphisolution}),
with momentum along $y$, transverse to the direction of action of the
trapping potential.

\section{ Discrete fermion spectrum}\label{boundstateonk}

In this section, we find the fermion bound states by solving the Schr\"{o}%
dinger equation on the background of the kink from the $\phi ^{4}$ theory
given by Eq. (\ref{kinkphisolution}). In this case, we follow closely the
approach developed by Chu and Vachaspati \cite{ChuPRD2008}. For the clarity
of the presentation and to define our notation, we repeat the essential
steps of that work for the 2+1 case, including the Lorentz symmetry breaking
terms of the model.

To begin, we rewrite the potentials $V_{\pm }(\phi )$ using the kink
solution Eq. (\ref{kinkphisolution}). In this case, we get%
\begin{equation}
V_{\pm }(z)=G^{2}-G(G\pm \alpha _{0})\mathrm{sech}^{2}\left( \alpha
_{0}z/(\hbar v_{F})\right) ,  \label{kpot1}
\end{equation}%
where $G=g\eta $ and $\alpha _{0}$ is given by Eq. (\ref{alpha0}). The $%
\hbar v_{F}$ factor gives $z$ in units of distance.  Next, one needs to calibrate the 
model parameters to a real situation, as we explain in the following.

In a recent bilayer graphene experiment, it was obtained the images of conducting channels driven by 
topological states\cite{YinNATC16}.  
The domain wall was created by deformations of one of the layers. The observed conducting channels correspond to states
localized at the edges of the domain wall. This experimental configuration leads to the formation of a 
topological zero mode states which 
is localized at the border of the domain wall. Naively we can expect the destruction of 
the peak at the center of the 
zero mode, as the  potential created by the topological deformation for this experimental set up, 
should be deformed at the middle of the
 domain wall to include a small barrier at the center, which can imply a double humped state. Of course 
 this effect is not included in our analytical 
model of the kink, as the effective potential is symmetrical and has the deep at the center of the kink. 

\begin{figure}[h]
\includegraphics[scale=0.78]{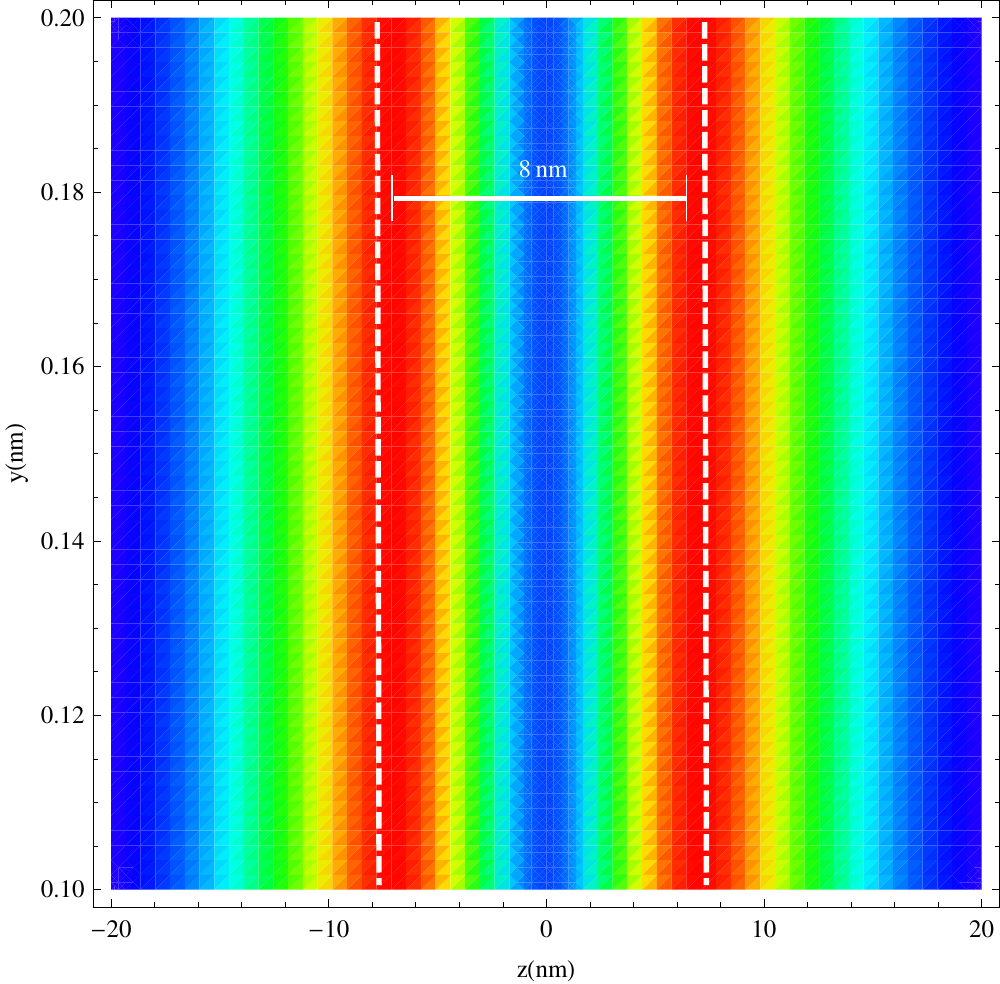}
\caption{Charge carrier density  distribution corresponding to the first excited state given in Fig. \ref{FIGspect}. }
\label{FIGdensdistr} 
\end{figure}

Nevertheless the first excited shows the general characteristic peaks at the edges of the domain wall, and due to that 
we set the parameters of the model to reproduce the relative distance between the two maximum of the first excited state to the observed
separation of the conducting channels of 8 nm. We obtain a kink
parameter of $\alpha _{0}=20\,\text{meV}$ and $G=500$ meV. As we are going to see the deep at the minimum of the excited state 
probability density is controlled by the ratio $G/\alpha_0$ which should be at least about $\sim 10$. To illustrate our choice of parameters, in Fig. \ref{FIGdensdistr} we anticipate our results for the charge carrier density of the first excited  trapped state, where it is clearly seen  the separation of 8 nm between the peaks, and they are located at edge of the domain wall.

We  will explore different cases of
 \begin{equation}\label{alpha0vphivf} 
 \alpha _{0}=1\,\text{meV}\,[v_{\phi }/v_{F}]^{-1}\, , 
\end{equation} 
 to check the effect of breaking the Lorentz 
 invariance, reminding that the Fermi velocity for the electron is $v_{F}=c/300$ in graphene. 
 
 We should emphasize that the kink in the present topological model, is much simpler than the alluded experimental set up. There are other interesting topological models of superposition of kinks   \cite{BazeiaPRD00,DutraPLB05,RafaelCQG11}, which could generate a potential closer to the experimental set up, however it is not yet known the
analytical solutions of the trapped fermionic modes in these non-trivial kink backgrounds.

It is worth noting that for values of $G>0$ the potential $V_{+}(z)$ has the
asymptotic maximum of $\lim_{z\rightarrow \pm \infty }V(z)=\eta ^{2}g^{2}$,
and minimum value of $-G \alpha _{0}$ at $z=0$. As it is well known from
quantum mechanics in one-dimension, a time-independent attractive potential
that tends to zero asymptotically will have at least one bound state. In
addition, the Schr\"{o}dinger-like equation derived for the upper component,
has a particular combination of the kink and its derivative, which gives a
zero mode \cite{RebbiPRD1976}. In general grounds, the zero mode solution of
the Dirac equation for a scalar domain wall presents a zero
mode (see e.g. \cite{VasBook2006}), and was theoretically discovered in the
graphene layer near the wall in \cite{SemPRL08}. Later on, in Ref. \cite%
{SasNJP10} the zero mode solution was found for the Dirac electron for
strained graphene nanoribbons in a soliton trap.

\begin{figure}[h]
 \includegraphics[scale=0.8]{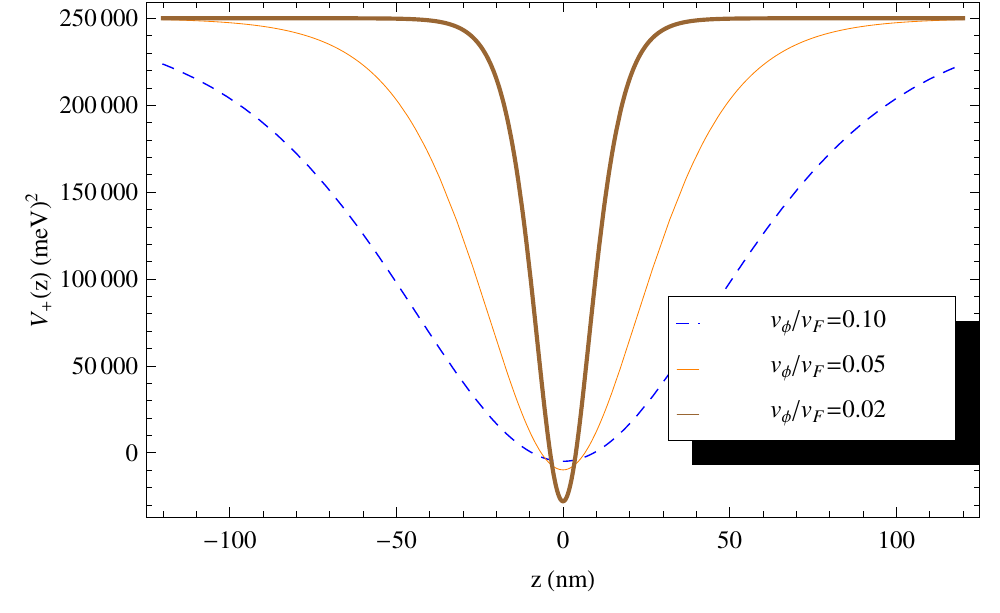} %
\includegraphics[scale=0.8]{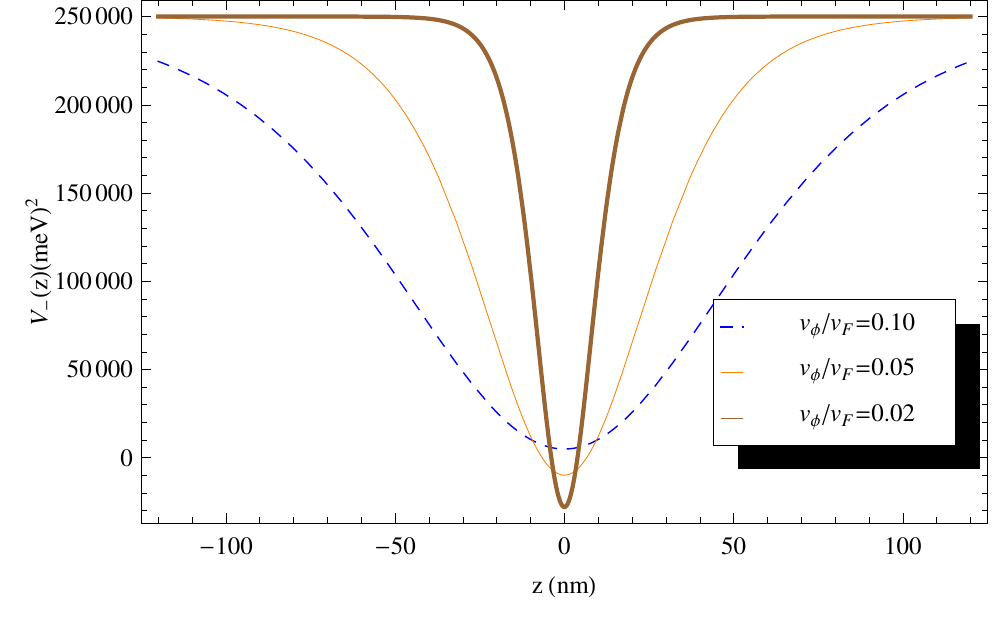}
\caption{ Effective potentials $V_{+}(z)$ (upper frame) and $V_{-}(z)$
(lower frame) obtained with $\sqrt{2\lambda }\eta =1$
meV, $G \,=500\,\text{\ meV}$ and $ \alpha _{0}=1\,\text{%
meV}\,[v_{\phi }/v_{F}]^{-1}$, for different ratios $v_{\phi %
}/v_{F}$ of $0.10$ (dashed line), $0.05$ (thin solid line) and $0.02$ (thick
solid line).}\label{FIGVP}
\end{figure}

Therefore, it is guaranteed that $V_{+}(z)$ has at least one bound state for
every $g$, which is the zero mode. Moreover, since $V_{+}(z)$ gets deeper
with increasing $g$, there are more and more bound states that appear with
larger values of $g$ \cite{ChuPRD2008}. The domain wall created by a kink in
2D materials can present a rich spectrum near the wall driven by the
effective potentials $V_\pm(z)$, illustrated in Fig. \ref{FIGVP}. In addition,
the Lorentz violating term drives the potentials by changing the relative
speed of the scalar particle to the Fermi velocity Eq. (\ref{velratio})
allowing to dial the spectrum with $v_\phi/v_F$. We observe in the upper panel of
 Fig. \ref{FIGVP} that
the increase of $v_\phi/v_F$ turns $V_+(z)$ shallow but wider, which indeed
increases the number of excited states, as can be seen in $V_-(z)$ in the lower panel 
of the figure, which becomes deeper and wider allowing more states. We are going to
detail analytically such properties of the spectrum.

In order to solve the Dirac equation, we need a non-trivial bound state of
the $\beta _{-}$ Schr\"{o}dinger-like equation which has the same energy
eigenvalue as for $\beta _{+}$. Only then $\beta _{\pm }$ will solve the
first order equations, Eq.~(\ref{betaequations}), except if $\varepsilon =0$
and for that we can take $\beta _{-}=0$. For $0<g\leq 1$, $V_{-}$ is in the
shape of a potential barrier and clearly has no bound states. This shows
that for $0<g\leq 1$, the only possible bound state is with $\varepsilon =0$
and ${\beta }_{-}=0$; the and the zero mode solution is 
\begin{equation}
\beta _{+}(z)\equiv \beta _{zm}(z)=\mathrm{sech}^{\vartheta _{0}}(\alpha
_{0}z/(\hbar v_{F})),  \label{betazm}
\end{equation}%
where $\vartheta _{0}=G /\alpha _{0}$. This zero mode solution near the wall
matches the one found in \cite{SemPRL08}. This state corresponds to a
massless particle moving in the transverse direction, $y$, with momentum $w$%
, and total energy $E$, as well.

Now, since that there are more bound states for $g>1$, we will begin our
search for these bound states. On the other hand, let us now carry out the
quantitative calculation of the stationary states $\beta _{\pm }(z)$, for do
this we write 
\begin{equation}
\beta _{\pm }(z)=\mathcal{N}_{\pm }\mathrm{sech}^{\vartheta }(\alpha
_{0}z/(\hbar v_{F}))F_{\pm }(z),  \label{q1}
\end{equation}%
where $\vartheta \equiv \left( \sqrt{G^{2}-\varepsilon ^{2}}\right) /\alpha
_{0}$. Then, substituting the Eq. (\ref{q1}) into Eqs. (\ref{schrodinger})
and (\ref{schrodinger1}), we obtain 
\begin{eqnarray}
&&\left. \partial _{z}^{2}F_{+}+2\alpha _{0}\vartheta \tanh (\alpha
_{0}z/(\hbar v_{F}))\partial _{z}F_{+}\right.  \label{q2.1} \\
&&\left. +\left[ \alpha _{0}^{2}\vartheta (1+\vartheta )-G(G+\alpha _{0})%
\right] \mathrm{sech}^{2}(\alpha _{0}z/(\hbar v_{F}))F_{+}(z)=0,\right. 
\notag \\
&&  \notag \\
&&\left. \partial _{z}^{2}F_{-}+2\alpha _{0}\vartheta \tanh (\alpha
_{0}z/(\hbar v_{F}))\partial _{z}F_{-}\right.  \label{q4.1} \\
&&\left. +\left[ \alpha _{0}^{2}\vartheta (1+\vartheta )-G(G-\alpha _{0})%
\right] \mathrm{sech}^{2}(\alpha _{0}z/(\hbar v_{F}))F_{-}(z)=0.\right. 
\notag
\end{eqnarray}

According to \cite{ChuPRD2008}, by the variable transformation 
\begin{equation}
u=\frac{1}{2}\,\left[ 1-\tanh \left( \alpha _{0}z/(\hbar v_{F})\right) %
\right] ,
\end{equation}%
the equations (\ref{q2.1}) and (\ref{q4.1}), becomes 
\begin{eqnarray}
&&\left. u(u-1)\partial _{u}^{2}F_{+}(u)+(\vartheta +1)(2u-1)\partial
_{u}F_{+}(u)+\right.  \notag \\
&&\left. \left[ \vartheta (\vartheta +1)-G(G+\alpha _{0})/\alpha _{0}^{2}%
\right] F_{\pm }(u)=0,\right.  \label{q5} \\
&&  \notag \\
&&\left. u(u-1)\partial _{u}^{2}F_{-}(u)+(\vartheta +1)(2u-1)\partial
_{u}F_{-}(u)+\right.  \notag \\
&&\left. \left[ \vartheta (\vartheta +1)-G(G-\alpha _{0})/\alpha _{0}^{2}%
\right] F_{-}(u)=0,\right.  \label{q7}
\end{eqnarray}

As we can see, the above equations are the so-called hypergeometric
equations, which has the following solutions%
\begin{eqnarray}
&&\left. F_{+}(u)=\mathcal{N}_{+}^{(1)}\mathcal{F}\left[ A_{+},B_{+};C_{+};u%
\right] \right. \\
&&\left. +\mathcal{N}_{+}^{(2)}\mathcal{F}\left[
A_{+}+1-C_{+},B_{+}+1-C_{+};2-C_{+};u\right] ,\right.  \notag  \label{q9}
\end{eqnarray}%
\begin{eqnarray}
&&\left. F_{-}(u)=\mathcal{N}_{-}^{(1)}\mathcal{F}\left[ A_{-},B_{-};C_{-};u%
\right] \right. \\
&&\left. +\mathcal{N}_{-}^{(2)}\mathcal{F}\left[
A_{-}+1-C_{-},B_{+}+1-C_{-};2-C_{-};u\right] ,\right.  \notag  \label{q12}
\end{eqnarray}%
\ where the arguments of the hypergeometric function are defined as 
\begin{eqnarray}
A_{\pm } &\equiv &\vartheta _{\pm }+\frac{1}{2}\, \mp \frac{1}{2} -\frac{G}{%
\alpha _{0}}, \\
B_{\pm } &\equiv &\vartheta _{\pm }+\frac{1}{2}\, \pm \frac{1}{2} +\frac{G}{%
\alpha _{0}} , \\
C_{\pm } &\equiv &\vartheta _{\pm }+1.
\end{eqnarray}

Therefore, we can write the solutions for $\beta _{\pm }(z)$ in the
following form 
\begin{eqnarray}
&&\left. \beta _{\pm }(z)=\mathcal{N}_{\pm }^{(1)}\mathrm{sech}^{\vartheta
}(\alpha _{0}z/(\hbar v_{F}))~\mathcal{F}[A_{\pm },B_{\pm };C_{\pm
};u]\right.  \label{q151} \\
&&\left. +\mathcal{N}_{\pm }^{(2)}e^{\alpha _{0}\vartheta z}\mathcal{F}%
[A_{\pm }-C_{\pm }+1,B_{\pm }-C_{\pm }+1;2-C_{\pm };u].\right.  \notag
\end{eqnarray}

As shown in \cite{ChuPRD2008}, for reason of normalizability, it is
necessary to impose that $\mathcal{N}_{\pm }^{(2)}=0$. Thus, the equation (%
\ref{q151}) becomes%
\begin{equation}
\beta _{\pm }(z)=\mathcal{N}_{\pm }^{(1)}\mathrm{sech}^{\vartheta }(\alpha
_{0}z/(\hbar v_{F}))~\mathcal{F}[A_{\pm },B_{\pm };C_{\pm };u].  \label{q16}
\end{equation}

In addition, we also have the condition 
\begin{equation}
\vartheta _{n}^{\pm }-\frac{G}{\alpha _{0}}+\frac{1}{2}\mp \frac{1}{2}%
=-n_{\pm }\in \mathbb{Z}^{-}.  \label{bnpm}
\end{equation}

Therefore, after straightforwardly mathematical manipulations, we obtain the
energy spectrum 
\begin{eqnarray}
&&E_{n_{-}}=\pm \sqrt{(\hbar \omega )^{2}+\alpha _{0}(n_{-}+1)\left[
2G-\alpha _{0}(n_{-}+1)\right] },  \notag \\
&&E_{n_{+}}=\pm \sqrt{(\hbar \omega )^{2}+n_{+}\alpha _{0}(2G-\alpha
_{0}n_{+})}\,.  \label{band1}
\end{eqnarray}%
At this point, it is important to remark that the energy eigenvalues
coincide $E_{n_{+}}=E_{n_{-}}\equiv E_{n}$, which implies in the consistence
condition $n_{+}-n_{-}=1$. Moreover, the normalizability requires $\vartheta
_{n}^{+}>0$, as a consequence we have the constraint 
\begin{equation}
1\leq n_{+}<2G/\alpha _{0}\,\,.  \label{numberexc}
\end{equation}%
The consistence condition for the quantum numbers leads to $\vartheta
_{n}^{\pm }\equiv \vartheta _{n}$, where $n\equiv n_{+}$. Just to remind the
zero mode is indexed with $n=0$.

Our solution reduces to the ones found in \cite{ChuPRD2008}, when $%
\alpha_0\to 1$, namely the LV term of the scalar field in the Lagrangian
vanishes. Note that as have shown the zero mode still survives,
independently of this LV interaction term. It builds a zero gap band, as
already found in previous works \cite{SemPRL08} and \cite{SasNJP10}.

\section{Discussion}

As we have already discussed, 
the motivation for the choice of the parameters 
comes from the experimental work \cite{YinNATC16}, where they found by
imaging the confined states that the associated conducting channels are
mainly located close to the two edges of the domain wall. In that experiment
the distance between the two peaks is about 8 nm. In our model the zero mode 
is peaked at the center of the domain wall, while the first excited state is concentrated
at the edges of the domain wall.

\subsection{Spectrum}

 In order to give some concrete scales of the spectrum, we find the 
 gap $\Delta$  between 
 the valence and conducting bands belonging to the first excited state in the
trap potential obtained with  $n=n_{+}=1$ and $\omega =0$ in Eq. (\ref%
{band1}): 
\begin{equation}
\Delta =2\sqrt{\alpha _{0}(2G-\alpha _{0})}\,,  \label{gap}
\end{equation}%
with  $\alpha _{0}=20\,\text{meV}$ and $G=500$ meV, one gets
$\Delta= 239$ meV, which is quite large when  compared to the value of 80 meV from \cite{YinNATC16}. This deference comes from the description of potential in the experimental situation, where one expects to have a  wider potential with  a barrier in the middle and therefore the level position can be
shifted down.

\begin{figure}[tbh]
\includegraphics[width=0.45\textwidth,left]{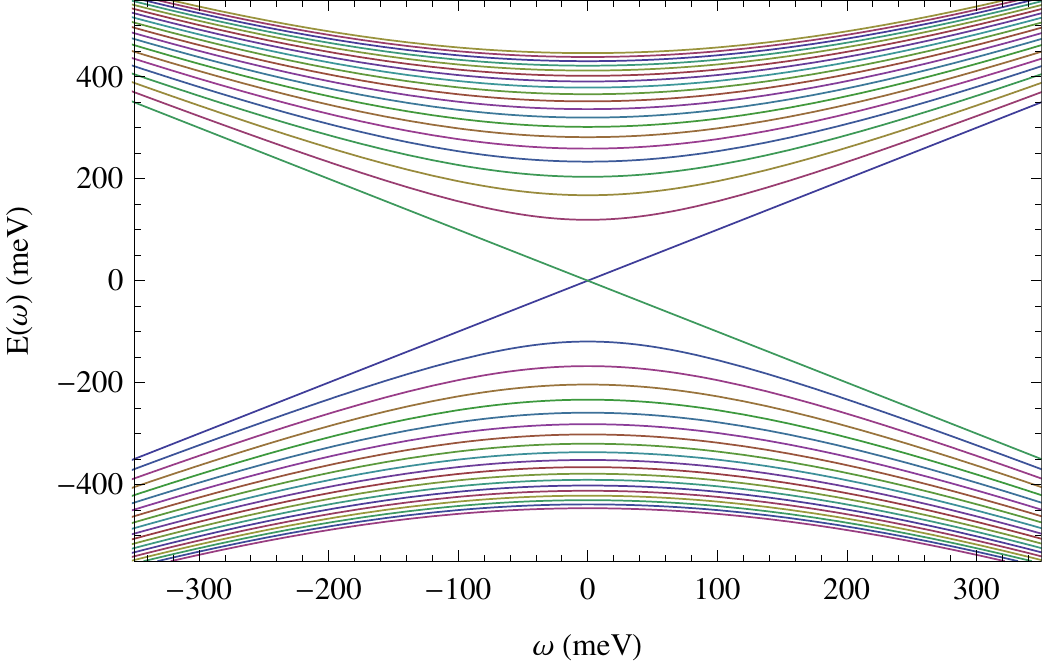}
\caption{ Band structure for the domain wall obtained with $G\,=500\,\text{meV}$ and $%
 \alpha _{0}=20\,\text{meV}$ for the first 39 states, including the zero mode. In this situation the number of states is 101 from $n_{max}=50$ as follows from
Eq. (\ref{numberexc}).}
\label{Fig3}
\end{figure}

The
spectrum for these parameters is shown in Fig. \ref{Fig3}. The dispersion relation
for the  zero mode and the first 38 excited states are presented. The chosen parameters leads to $n_{max}=50$ from Eq. (\ref{numberexc}), giving a large number of conducting channels, concentrated at the edges of the kink.

We observe that the quantity  $m^{\text{eff}}_n\equiv\sqrt{n\alpha_0(2G-\alpha_0\,n)}$ can
be interpreted as an effective mass for each of the excited trapped states,
which can be written in terms of the gap as: 
\begin{equation}
m^{\text{eff}}_n=\sqrt{n\,\left(\frac{\Delta^2}{4}-\alpha_0^2(n-1)\right)}\,
.
\end{equation}

\subsection{Charge carrier density distribution}

On the kink background the wave function of the n-th fermionic state, $%
\psi_n(y,z;t)$, trapped by the wall, for positive and negative energies,
including the zero mode solutions are built from Eqs. (\ref{sd2}), (\ref%
{betazm}) and (\ref{q16}).

The zero mode eigenstate of the Dirac Hamiltonian trapped by the soliton is: 
\begin{equation}
\left. \psi _{0}(y,z;t)=N_{0}e^{i\,\omega ({y/v_{F}}-\,t)}\,\,\mathrm{sech}%
^{\vartheta _{0}}\left( \frac{\alpha _{0}}{\hbar v_{F}}z\right) \,\left[ 
\begin{array}{c}
1 \\ 
1%
\end{array}%
\right] ,\right.
\end{equation}%
which is the solution already found and thoroughly discussed in \cite%
{SasNJP10}, apart the LV effect. We have to observe that the zero-mode state
in the layer corresponds to the solution of the one dimensional Dirac
equation with a "massless" fermion, as the dispersion relation is $E=\hbar\omega $%
. In addition the spinor state is an eigenstate of $\gamma ^{0}\gamma ^{2}$.
The trapping potential gives a width to the charge distribution along the $z$
direction, as carried the $\mathrm{sech}$ term.

\begin{figure}[thb]
 \includegraphics[scale=0.85]{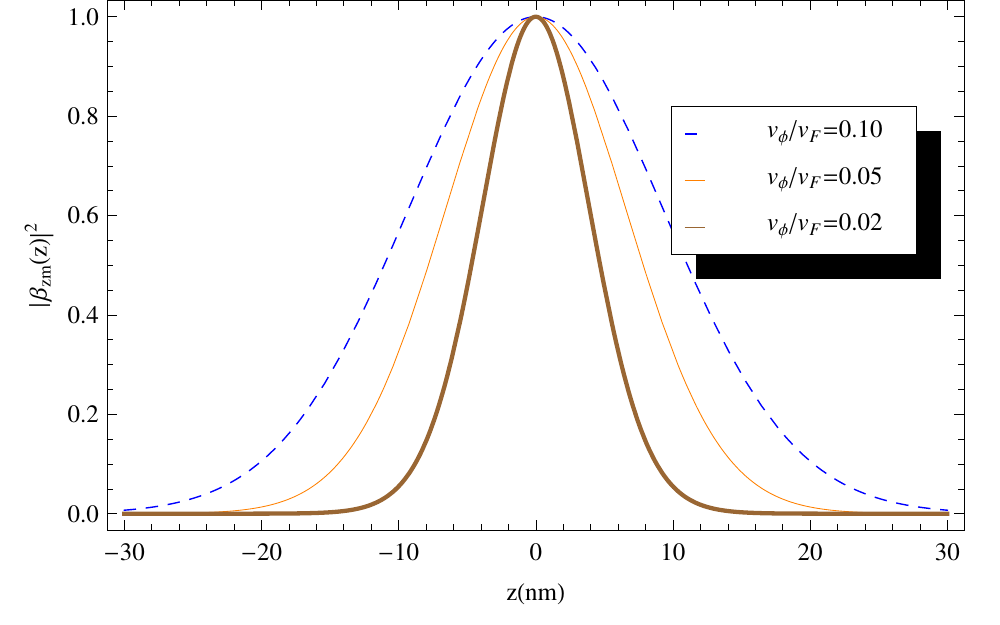}
\caption{Density profile of the zero-mode solution with LV term obtained
with $G\,=500\,\text {meV}$ and $ \alpha_0= 1\, \text{meV} \,[v_ \phi/v_F]^{-1}$,
for different ratios $v_ \phi/v_F$ of $0.10$ (dashed line), $0.05$
(thin solid line) and $0.02$ (thick solid line). }\label{FIG:ZM}
\end{figure}

The LV term in the scalar field Lagrangian changes only the velocity of the
scalar particle $(v_\phi/v_F=\sqrt{1-k^{11}})$, and in the zero mode
solution it only modifies the $\alpha_0$ and the net effect is to broaden or
shrink the wave function, as one sees in Fig. \ref{FIG:ZM} for $v_\phi/v_F$ equals to
0.1, 0.05 and 0.02; as the velocity ratio becomes larger wider is the state
density, as the effective potential also becomes wider as shown in Fig. \ref{FIGVP}.
That comes because the value of $\alpha_0$ according to Eq. (\ref{alpha0vphivf}), is inversely proportional to  $v_\phi/v_F$, and
$\alpha_0$ larger/smaller implies in confining potentials narrower/wider.

We have shown results for small values of the velocity ratio and of course
could be enhanced by changing more drastically the ratio. This effect could
be useful to dial the width of the charge density in the fermion zero mode
state in the solitonic trap by modifying the properties of the collective
mode propagation, either doping the 2D material, or introducing defects on
the lattice, or by a varying strain, etc. This one-dimensional conducting
state lives at the middle of the domain wall, while as we are going to show
the higher excited states are located mainly close to the edges of the kink.

\begin{figure}[tbh]
 \includegraphics[width=0.75\textwidth,center]{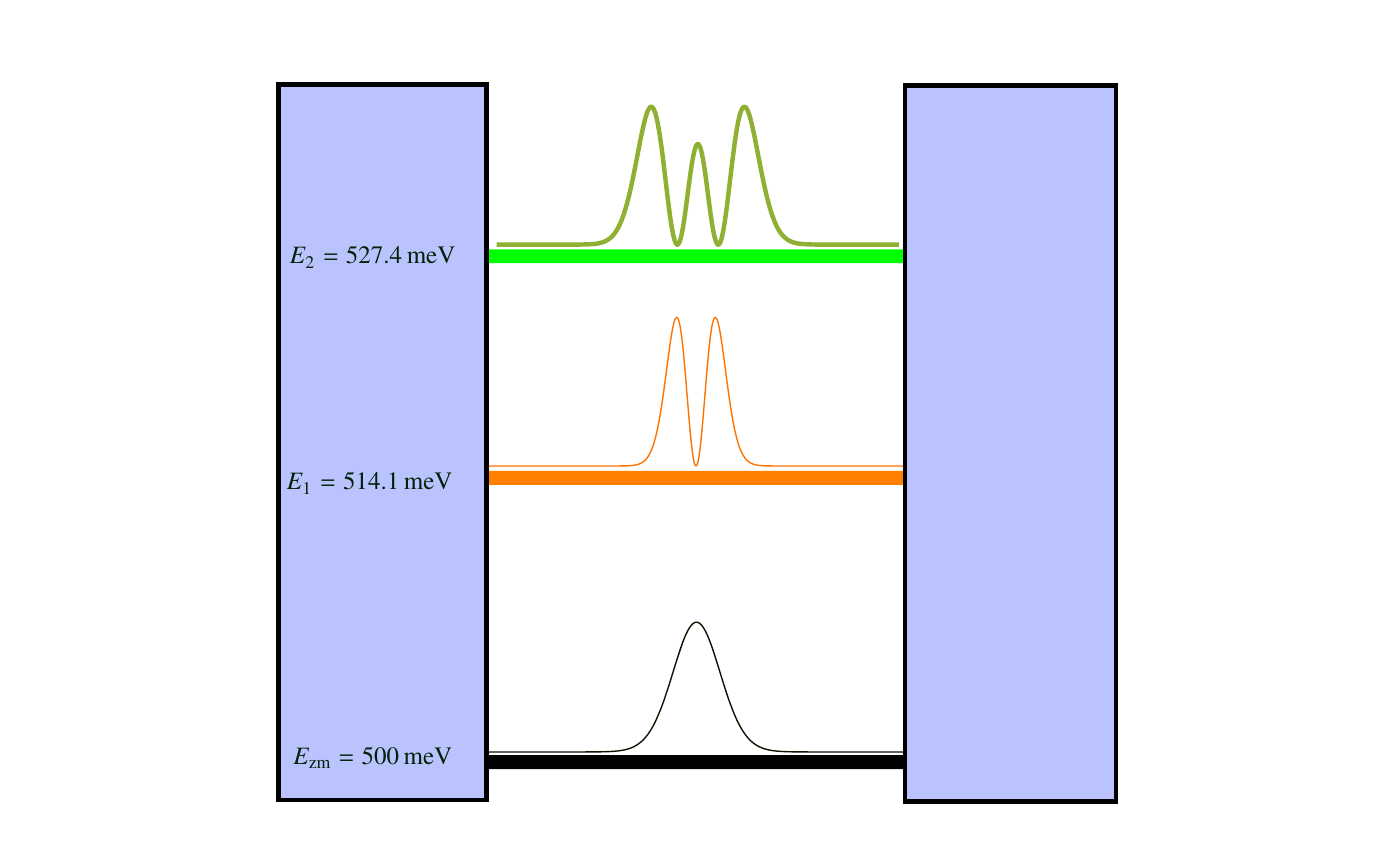}
\caption{Probability density of the bound states with positive energy. The
electron has a momentum $k_{y}= \omega $ along the wall, with $\hbar%
 \omega =500$ meV, $ \alpha _{0}=20$ meV and $G=500$ meV.} \label{FIGspect}
\end{figure}

The wave function of the excited states in the trapping potential of the
kink with $n\geq 1$, are given by: 
\begin{eqnarray}\label{wfexc}
&&\left. \psi _{n}(y,z;t)=N_{n}e^{i(\omega y-E_{n}\,t)}\mathrm{sech}%
^{\vartheta _{n}}\left( \alpha _{0}z/(\hbar v_{F})\right) \times \right. 
\notag \\
&&\left. \left[ 
\begin{array}{c}
-E_{n}F_{+}(z)-\alpha _{0}nF_{-}(z) \\ 
-E_{n}F_{+}(z)+\alpha _{0}nF_{-}(z)%
\end{array}%
\right] ,\right.
\end{eqnarray}%
where%
\begin{equation}
\vartheta _{n}=\frac{G}{\alpha _{0}}-n  \label{bnpm-1}
\end{equation}%
\begin{eqnarray}
F_{+}(z) &=&\mathcal{F}\left[ A_{+},B_{+};C_{+};\frac{1}{2}\,\left[ 1-\tanh
\left( \alpha _{0}z/(\hbar v_{F})\right) \right] \right] ,  \notag \\
&& \\
F_{-}(z) &=&\mathcal{F}\left[ A_{-},B_{-};C_{-};\frac{1}{2}\,\left[ 1-\tanh
\left( \alpha _{0}z/(\hbar v_{F})\right) \right] \right] .  \notag \\
&&
\end{eqnarray}

The effect of the Lorentz symmetry breaking term in the scalar field
Lagrangian, which corresponds to the scaling of the propagation velocity by $%
\alpha _{0}^{-1}$ or $\sqrt{1-k^{11}}$, allows to increase/decrease the
number of trapped states by increasing/decreasing the speed of the scalar
mode in the 2D material, as the condition for the number of states (\ref%
{numberexc}) shows for a fixed value of $G$. 
Furthermore, the decrease in $\alpha_0$ increase  the density of states.

In Fig. \ref{FIGspect}, we illustrate the charge carrier density for the lowest states obtained
 with $\hbar\omega= 500$ meV, $\alpha_0=20$ meV and $G=$ 500 meV.
In this case the density is dominated by $F_+(z)$ due to the energy factor, as seen in
Eq.(\ref{wfexc}). The effect of the LV term expressing the 
change in the relative speeds of the electron and the scalar quantum, modifies $\alpha_0$ as shown 
by Eq. (\ref{alpha0vphivf}), impact both  the trapped excited spectrum and in the associated charge carrier densities. For example, by decreasing $\alpha_0$ from the increase of the scalar mode velocity with respect to $v_F$, the domain wall becomes softer  the charge carrier density swallows.

\section{Conclusions}

\label{summary}

In this work we have investigated the full spectrum of two-dimensional
fermion states in a scalar soliton trap with a Lorentz breaking background.
It is important to remark that in the context of the novel 2D materials, 
the Lorentz symmetry should not be strictly valid for a field theory 
that aims to represent the charge carriers
close to the Dirac points and the collective modes of the lattice represented by
bosonic fields. The Dirac electron and  bosonic fields reflect the 
non-relativistic dynamics of the 
electrons and lattice. In that sense, it is natural that the bosonic fields 
have propagation velocities different from $v_F$, and the effect of this kind 
of Lorentz violating effect should be expected when modelling the electronic and lattice excitations in 2D materials, like 
graphene. Therefore, it is necessary that the Lorentz symmetry breaking should be 
taken into account in the chosen field theory model. Having that motivation,
we have performed an analysis of this particular Lorentz violation  effect in a 2D material with a relativistic model of the electrons and fields. 

We explore theoretically, the asymmetry  between the  Fermi  and scalar 
mode velocities, in the wave function of the trapped electrons in a kink background.
To accomplish that, we extended the analytical methods developed in the context of 1+1 
field theories to study the role of confined fermions to the stability of cosmic strings 
(see \cite{ChuPRD2008}),  to explore the effect of the Lorentz symmetry breaking in the charge carrier density of 2D materials in the presence of a domain wall with a 
kink profile. 
The width and the depth of the trapping potential from the kink in the presence of the 
Lorentz violating effect associated with the asymmetry of the Fermi and scalar mode 
velocities can be manipulated, and is reflected in the band structure and 
properties of the levels localized along the kink axis and  propagating in the transverse direction. 
The present model provides an analytical form for the rich spectrum 
enlarging a previous findings  of the edge states obtained with 
domain wall \cite{SemPRL08}  and in particular in strained graphene nanoribbon
in a chiral gauge theory \cite{SasNJP10}. In future it will be interesting to generalize our analysis, when transitions between the two valleys are allowed in the field theoretical model, which leads a four dimensional representation of the spinor.

In summary, the  bound states solutions of Dirac equation in the kink background 
are not destroyed by the Lorentz violation due to the different velocities of the Dirac electron and boson modes. One should also expect that theories with gauge fields
that aims to describe  electronic properties of 2D materials are not necessarily covariant,
and Lorentz symmetry breaking effects should be taken into account. We found that
by tuning  the Lorentz breaking parameter,
 it is possible to change the density of trapped states and make the associated charge 
carrier density  wider or narrow, which may be of interest in actual  
applications of 2D materials to electronic devices. Reversely, one could also search the consequences of Lorentz violation in controlled table top experiments with 2D materials, which could be relevant to understand subtle phenomena in the universe.

\begin{acknowledgments}
RACC would like to thank S\~{a}o Paulo Research Foundation (FAPESP), grant
2016/03276-5, for financial support. WDP, ASD and TF thank to CAPES, CNPq
and Fapesp for financial support. RACC gratefully acknowledge Jos\'{e}
Abdalla Helayel-Neto for discussions at an early stage.
\end{acknowledgments}

\end{document}